\documentstyle[12pt,epsf]{article}
\catcode`@=11
\def\chkspace{%
  \relax   
  \begingroup\ifhmode\aftergroup\dochksp@ce\fi\endgroup}
\def\dochksp@ce{%
  \unskip              
  \futurelet\chkspct@k\d@chkspc  
}
\def\d@chkspc{%
  \let\nxtsp@ce=\relax
  \ifx\chkspct@k.\else     
    \ifx\chkspct@k,\else
      \ifx\chkspct@k;\else
        \ifx\chkspct@k!\else
          \ifx\chkspct@k?\else
            \ifx\chkspct@k:\else
              \ifx\chkspct@k)\else
              \ifx\chkspct@k(\else
                \ifx\chkspct@k]\else
                  \ifx\chkspct@k-\else
                    \ifx\chkspct@k\egroup\else  
                      \let\nxtsp@ce=\put@space  
                    \fi
                  \fi
                \fi
              \fi
              \fi
            \fi
          \fi
        \fi
      \fi
    \fi
  \fi
  \nxtsp@ce
}
\def\put@space{$\;$}
\catcode`@=12

\def\ra{{$\rightarrow$}\chkspace}
\def\etal{{\it et al.}\chkspace}

\def\ie{{\it i.e.}\chkspace}

\def\eg{{\it eg.}\chkspace}

\def\apriori{{\it a priori}\chkspace}

\def\ep{{e$^+$e$^-$}\chkspace}
\def\epa{{e$^+$e$^-$ annihilation}\chkspace}

\def\gluino{\relax\ifmmode \tilde{g} \else $\tilde{g}$ \fi\chkspace}

\def\qq{\relax\ifmmode q\overline{q}
\else $q\overline{q}$ \fi\chkspace}

\def\bb{\relax\ifmmode b\bar{b}
       \else $b\bar{b}$ \fi\chkspace}
\def\ccrm{\relax\ifmmode {\rm c}\bar{\rm c}
       \else ${\rm c}\bar{\rm c}$ \fi\chkspace}
\def\cc{$c\bar{c}$ \chkspace}
\def\tt{\relax\ifmmode {\rm t}\bar{\rm t}
       \else ${\rm t}\bar{\rm t}$ \fi\chkspace}
\def\ss{\relax\ifmmode {\rm s}\bar{\rm s}
       \else ${\rm s}\bar{\rm s}$ \fi\chkspace}
\def\uu{\relax\ifmmode {\rm u}\bar{\rm u}
       \else ${\rm u}\bar{\rm u}$ \fi\chkspace}
\def\dd{\relax\ifmmode {\rm d}\bar{\rm d}
       \else ${\rm d}\bar{\rm d}$ \fi\chkspace}

\def\qqg{\relax\ifmmode q\overline{q}g
\else $q\overline{q}g$ \fi\chkspace}

\def\bbg{$b\overline{b}g$\chkspace}

\def\afb{\relax\ifmmode A_{FB} \else
{{$A_{FB}$}}\fi\chkspace}
\def\afbb{\relax\ifmmode A_{FB}^b \else
{{$A_{FB}^b$}}\fi\chkspace}
\def\pafb{\relax\ifmmode \tilde{A}_{FB} \else
{{$\tilde{A}_{FB}$}}\fi\chkspace}
\def\pafbb{\relax\ifmmode \tilde{A}_{FB}^b \else
{{$\tilde{A}_{FB}^b$}}\fi\chkspace}

\def\pafbzo{\relax\ifmmode \tilde{A}_{FB}|_{O(0)} \else
{{$\tilde{A}_{FB}|_{O(0)}$}}\fi\chkspace}
\def\pafbfo{\relax\ifmmode \tilde{A}_{FB}|_{\oalp} \else
{{$\tilde{A}_{FB}|_{\oalp}$}}\fi\chkspace}
\def\pafbso{\relax\ifmmode \tilde{A}_{FB}|_{\oalpsq} \else
{{$\tilde{A}_{FB}|_{\oalpsq}$}}\fi\chkspace}
\def\pafbto{\relax\ifmmode \tilde{A}_{FB}|_{\oalpc} \else
{{$\tilde{A}_{FB}|_{\oalpc}$}}\fi\chkspace}

\def\pafbbzo{\relax\ifmmode \tilde{A}_{FB}^b|_{O(0)} \else
{{$\tilde{A}_{FB}^b|_{O(0)}$}}\fi\chkspace}
\def\pafbbfo{\relax\ifmmode \tilde{A}_{FB}^b|_{\oalp} \else
{{$\tilde{A}_{FB}^b|_{\oalp}$}}\fi\chkspace}
\def\pafbbso{\relax\ifmmode \tilde{A}_{FB}^b|_{\oalpsq} \else
{{$\tilde{A}_{FB}^b|_{\oalpsq}$}}\fi\chkspace}
\def\pafbbto{\relax\ifmmode \tilde{A}_{FB}^b|_{\oalpc} \else
{{$\tilde{A}_{FB}^b|_{\oalpc}$}}\fi\chkspace}

\def\afbo0{\tilde{A}_{FB}|_{O(0)}}
\def\afbo1{\tilde{A}_{FB}|_{\oalp}}
\def\afbo2{\tilde{A}_{FB}|_{\oalpsq}}
\def\afbo3{\tilde{A}_{FB}|_{\oalpc}}

\def\lam{\relax\ifmmode \Lambda_{\overline{MS}}
       \else {{$\Lambda_{\overline{MS}}$}}\fi\chkspace}
\def\lamuds{\relax\ifmmode \Lambda^{(3)}_{\overline{MS}}
       \else {{$\Lambda^{(3)}_{\overline{MS}}$}}\fi\chkspace}
\def\lamudsc{\relax\ifmmode \Lambda^{(4)}_{\overline{MS}}
       \else $\Lambda^{(4)}_{\overline{MS}}$\fi\chkspace}
\def\lamudscb{\relax\ifmmode \Lambda^{(5)}_{\overline{MS}}
       \else $\Lambda^{(5)}_{\overline{MS}}$\fi\chkspace}
\def\alpb{$\alpha_s^{b}$\chkspace}
\def\alpc{$\alpha_s^{c}$\chkspace}

\def\alpuds{$\alpha_s^{uds}$\chkspace}

\def\alp{\relax\ifmmode \alpha_s\else $\alpha_s$\fi\chkspace}
\def\alpbar{\relax\ifmmode \bar{\alpha_s}
       \else $\bar{\alpha_s}$\fi\chkspace}
\def\alpmz{\relax\ifmmode \alpha_s(M_Z)\else $\alpha_s(M_Z)$\fi\chkspace}
\def\alpmzsq{\relax\ifmmode \alpha_s(M_Z^2)
       \else $\alpha_s(M_Z^2)$\fi\chkspace}

\def\oalp{\relax\ifmmode O(\alpha_s)\else{{O($\alpha_s$)}}\fi\chkspace}
\def\oalpsq{\relax\ifmmode O(\alpha_s^2)
           \else{{O($\alpha_s^2$)}}\fi\chkspace}
\def\oalpc{\relax\ifmmode O(\alpha_s^3)
           \else{{O($\alpha_s^3$)}}\fi\chkspace}
\def\oalpf{\relax\ifmmode O(\alpha_s^4)
           \else{{O($\alpha_s^4$)}}\fi\chkspace}

\def\rb{\relax\ifmmode R_3^b/R_3^{all}
           \else{{$R_3^b/R_3^{all}$}}\fi\chkspace}
\def\rc{\relax\ifmmode R_3^c/R_3^{all}
           \else{{$R_3^c/R_3^{all}$}}\fi\chkspace}
\def\ruds{\relax\ifmmode R_3^{uds}/R_3^{all}
           \else{{$R_3^{uds}/R_3^{all}$}}\fi\chkspace}
\def\ri{\relax\ifmmode R_3^i/R_3^{all}
           \else{{$R_3^i/R_3^{all}$}}\fi\chkspace}
\def\rj{\relax\ifmmode R_3^j/R_3^{all}
           \else{{$R_3^j/R_3^{all}$}}\fi\chkspace}
\def\alpi{\relax\ifmmode \alpha^i_s/\alpha^{all}_s
           \else{{$\alpha^i_s/\alpha^{all}_s$}}\fi\chkspace}

\def\mbz{\relax\ifmmode m_b(M_Z)
           \else{{$m_b(M_Z)$}}\fi\chkspace}
\def\mbb{\relax\ifmmode m_b(M_b)
           \else{{$m_b(M_b)$}}\fi\chkspace}

\def\plb{Phys. Lett.\chkspace}
\def\npb{Nucl. Phys.\chkspace}

\def\prl{Phys. Rev. Lett.\chkspace}
\def\prd{Phys. Rev.\chkspace}

\def\z0{{$Z^0$}\chkspace}
\def\Dst{\relax\ifmmode {\rm D}^* \else {D$^*$}\fi\chkspace}
\def\Dpl{\relax\ifmmode {\rm D}^+ \else {D$^+$}\fi\chkspace}
\def\D0{\relax\ifmmode {\rm D}^0 \else {D$^0$}\fi\chkspace}
\def\Kst{\relax\ifmmode {\rm K}^* \else {K$^*$}\fi\chkspace}
\def\K0{\relax\ifmmode {\rm K}^0_s \else {K$^0_s$}\fi\chkspace}
\def\Kpl{\relax\ifmmode {\rm K}^+ \else {K$^+$}\fi\chkspace}
\def\Kstz{\relax\ifmmode {\rm K}^{*0} \else {K$^{*0}$}\fi\chkspace}
\def\PsfigVersion{1.9}
\ifx\undefined\psfig\else\endinput\fi
 
\let\LaTeXAtSign=\@
\let\@=\relax
\edef\psfigRestoreAt{\catcode`\@=\number\catcode`@\relax}
\catcode`\@=11\relax
\newwrite\@unused
\def\ps@typeout#1{{\let\protect\string\immediate\write\@unused{#1}}}
\ps@typeout{psfig/tex \PsfigVersion}
 
\def\figurepath{./}

\def\@nnil{\@nil}
\def\@empty{}
\def\@psdonoop#1\@@#2#3{}
\def\@psdo#1:=#2\do#3{\edef\@psdotmp{#2}\ifx\@psdotmp\@empty \else
    \expandafter\@psdoloop#2,\@nil,\@nil\@@#1{#3}\fi}
\def\@psdoloop#1,#2,#3\@@#4#5{\def#4{#1}\ifx #4\@nnil \else
       #5\def#4{#2}\ifx #4\@nnil \else#5\@ipsdoloop #3\@@#4{#5}\fi\fi}
\def\@ipsdoloop#1,#2\@@#3#4{\def#3{#1}\ifx #3\@nnil
       \let\@nextwhile=\@psdonoop \else
      #4\relax\let\@nextwhile=\@ipsdoloop\fi\@nextwhile#2\@@#3{#4}}
\def\@tpsdo#1:=#2\do#3{\xdef\@psdotmp{#2}\ifx\@psdotmp\@empty \else
    \@tpsdoloop#2\@nil\@nil\@@#1{#3}\fi}
\def\@tpsdoloop#1#2\@@#3#4{\def#3{#1}\ifx #3\@nnil
       \let\@nextwhile=\@psdonoop \else
      #4\relax\let\@nextwhile=\@tpsdoloop\fi\@nextwhile#2\@@#3{#4}}
\ifx\undefined\fbox
\newdimen\fboxrule
\newdimen\fboxsep
\newdimen\ps@tempdima
\newbox\ps@tempboxa
\long\def\fbox#1{\leavevmode\setbox\ps@tempboxa\hbox{#1}\ps@tempdima\fboxrule
    \advance\ps@tempdima \fboxsep \advance\ps@tempdima \dp\ps@tempboxa
   \hbox{\lower \ps@tempdima\hbox
  {\vbox{\hrule height \fboxrule
          \hbox{\vrule width \fboxrule \hskip\fboxsep
          \vbox{\vskip\fboxsep \box\ps@tempboxa\vskip\fboxsep}\hskip
                 \fboxsep\vrule width \fboxrule}
                 \hrule height \fboxrule}}}}
\fi
\newread\ps@stream
\newif\ifnot@eof       
\newif\if@noisy        
\newif\if@atend        
\newif\if@psfile       
{\catcode`\%=12\global\gdef\epsf@start{
\def\epsf@PS{PS}
\def\epsf@getbb#1{%
\openin\ps@stream=#1
\ifeof\ps@stream\ps@typeout{Error, File #1 not found}\else
   {\not@eoftrue \chardef\other=12
    \def\do##1{\catcode`##1=\other}\dospecials \catcode`\ =10
    \loop
       \if@psfile
	  \read\ps@stream to \epsf@fileline
       \else{
	  \obeyspaces
          \read\ps@stream to \epsf@tmp\global\let\epsf@fileline\epsf@tmp}
       \fi
       \ifeof\ps@stream\not@eoffalse\else
       \if@psfile\else
       \expandafter\epsf@test\epsf@fileline:. \\%
       \fi
          \expandafter\epsf@aux\epsf@fileline:. \\%
       \fi
   \ifnot@eof\repeat
   }\closein\ps@stream\fi}%
\long\def\epsf@test#1#2#3:#4\\{\def\epsf@testit{#1#2}
			\ifx\epsf@testit\epsf@start\else
\ps@typeout{Warning! File does not start with `\epsf@start'.  It may not be a
 PostScript file.}
			\fi
			\@psfiletrue} 
{\catcode`\%=12\global\let\epsf@percent=
\long\def\epsf@aux#1#2:#3\\{\ifx#1\epsf@percent
   \def\epsf@testit{#2}\ifx\epsf@testit\epsf@bblit
	\@atendfalse
        \epsf@atend #3 . \\%
	\if@atend
	   \if@verbose{
		\ps@typeout{psfig: found `(atend)'; continuing search}
	   }\fi
        \else
        \epsf@grab #3 . . . \\%
        \not@eoffalse
        \global\no@bbfalse
        \fi
   \fi\fi}%
\def\epsf@grab #1 #2 #3 #4 #5\\{%
   \global\def\epsf@llx{#1}\ifx\epsf@llx\empty
      \epsf@grab #2 #3 #4 #5 .\\\else
   \global\def\epsf@lly{#2}%
   \global\def\epsf@urx{#3}\global\def\epsf@ury{#4}\fi}%
\def\epsf@atendlit{(atend)}
\def\epsf@atend #1 #2 #3\\{%
   \def\epsf@tmp{#1}\ifx\epsf@tmp\empty
      \epsf@atend #2 #3 .\\\else
   \ifx\epsf@tmp\epsf@atendlit\@atendtrue\fi\fi}

\chardef\psletter = 11 
\chardef\other = 12
 
\newif \ifdebug 
\newif\ifc@mpute 
\c@mputetrue 
 
\let\then = \relax
\def\r@dian{pt }
\let\r@dians = \r@dian
\let\dimensionless@nit = \r@dian
\let\dimensionless@nits = \dimensionless@nit
\def\internal@nit{sp }
\let\internal@nits = \internal@nit
\newif\ifstillc@nverging
\def \Mess@ge #1{\ifdebug \then \message {#1} \fi}
 
{ 
	\catcode `\@ = \psletter
	\gdef \nodimen {\expandafter \n@dimen \the \dimen}
	\gdef \term #1 #2 #3%
	       {\edef \t@ {\the #1}
		\edef \t@@ {\expandafter \n@dimen \the #2\r@dian}%
		\t@rm {\t@} {\t@@} {#3}%
	       }
	\gdef \t@rm #1 #2 #3%
	       {{%
		\count 0 = 0
		\dimen 0 = 1 \dimensionless@nit
		\dimen 2 = #2\relax
		\Mess@ge {Calculating term #1 of \nodimen 2}%
		\loop
		\ifnum	\count 0 < #1
		\then	\advance \count 0 by 1
			\Mess@ge {Iteration \the \count 0 \space}%
			\Multiply \dimen 0 by {\dimen 2}%
			\Mess@ge {After multiplication, term = \nodimen 0}%
			\Divide \dimen 0 by {\count 0}%
			\Mess@ge {After division, term = \nodimen 0}%
		\repeat
		\Mess@ge {Final value for term #1 of
				\nodimen 2 \space is \nodimen 0}%
		\xdef \Term {#3 = \nodimen 0 \r@dians}%
		\aftergroup \Term
	       }}
	\catcode `\p = \other
	\catcode `\t = \other
	\gdef \n@dimen #1pt{#1} 
}
 
\def \Divide #1by #2{\divide #1 by #2} 
 
\def \Multiply #1by #2
       {{
	\count 0 = #1\relax
	\count 2 = #2\relax
	\count 4 = 65536
	\Mess@ge {Before scaling, count 0 = \the \count 0 \space and
			count 2 = \the \count 2}%
	\ifnum	\count 0 > 32767 
	\then	\divide \count 0 by 4
		\divide \count 4 by 4
	\else	\ifnum	\count 0 < -32767
		\then	\divide \count 0 by 4
			\divide \count 4 by 4
		\else
		\fi
	\fi
	\ifnum	\count 2 > 32767 
	\then	\divide \count 2 by 4
		\divide \count 4 by 4
	\else	\ifnum	\count 2 < -32767
		\then	\divide \count 2 by 4
			\divide \count 4 by 4
		\else
		\fi
	\fi
	\multiply \count 0 by \count 2
	\divide \count 0 by \count 4
	\xdef \product {#1 = \the \count 0 \internal@nits}%
	\aftergroup \product
       }}
 
\def\r@duce{\ifdim\dimen0 > 90\r@dian \then   
		\multiply\dimen0 by -1
		\advance\dimen0 by 180\r@dian
		\r@duce
	    \else \ifdim\dimen0 < -90\r@dian \then  
		\advance\dimen0 by 360\r@dian
		\r@duce
		\fi
	    \fi}
 
\def\Sine#1%
       {{%
	\dimen 0 = #1 \r@dian
	\r@duce
	\ifdim\dimen0 = -90\r@dian \then
	   \dimen4 = -1\r@dian
	   \c@mputefalse
	\fi
	\ifdim\dimen0 = 90\r@dian \then
	   \dimen4 = 1\r@dian
	   \c@mputefalse
	\fi
	\ifdim\dimen0 = 0\r@dian \then
	   \dimen4 = 0\r@dian
	   \c@mputefalse
	\fi
	\ifc@mpute \then
		\divide\dimen0 by 180
		\dimen0=3.141592654\dimen0
		\dimen 2 = 3.1415926535897963\r@dian 
		\divide\dimen 2 by 2 
		\Mess@ge {Sin: calculating Sin of \nodimen 0}%
		\count 0 = 1 
		\dimen 2 = 1 \r@dian 
		\dimen 4 = 0 \r@dian 
		\loop
			\ifnum	\dimen 2 = 0 
			\then	\stillc@nvergingfalse
			\else	\stillc@nvergingtrue
			\fi
			\ifstillc@nverging 
			\then	\term {\count 0} {\dimen 0} {\dimen 2}%
				\advance \count 0 by 2
				\count 2 = \count 0
				\divide \count 2 by 2
				\ifodd	\count 2 
				\then	\advance \dimen 4 by \dimen 2
				\else	\advance \dimen 4 by -\dimen 2
				\fi
		\repeat
	\fi
			\xdef \sine {\nodimen 4}%
       }}
 
\def\Cosine#1{\ifx\sine\UnDefined\edef\Savesine{\relax}\else
		             \edef\Savesine{\sine}\fi
	{\dimen0=#1\r@dian\advance\dimen0 by 90\r@dian
	 \Sine{\nodimen 0}
	 \xdef\cosine{\sine}
	 \xdef\sine{\Savesine}}}
 
\def\psdraft{
	\def\@psdraft{0}
}
\def\psfull{
	\def\@psdraft{100}
}
 
\psfull
 
\newif\if@scalefirst
\def\psscalefirst{\@scalefirsttrue}
\def\psrotatefirst{\@scalefirstfalse}
\psrotatefirst
 
\newif\if@draftbox
\def\psnodraftbox{
	\@draftboxfalse
}
\def\psdraftbox{
	\@draftboxtrue
}
\@draftboxtrue
 
\newif\if@prologfile
\newif\if@postlogfile
\def\pssilent{
	\@noisyfalse
}
\def\psnoisy{
	\@noisytrue
}
\psnoisy
\newif\if@bbllx
\newif\if@bblly
\newif\if@bburx
\newif\if@bbury
\newif\if@height
\newif\if@width
\newif\if@rheight
\newif\if@rwidth
\newif\if@angle
\newif\if@clip
\newif\if@verbose
\def\@p@@sclip#1{\@cliptrue}

\newif\if@decmpr
 
\def\@p@@sfigure#1{\def\@p@sfile{null}\def\@p@sbbfile{null}
	        \openin1=#1.bb
		\ifeof1\closein1
	        	\openin1=\figurepath#1.bb
			\ifeof1\closein1
			        \openin1=#1
				\ifeof1\closein1%
				       \openin1=\figurepath#1
					\ifeof1
					   \ps@typeout{Error, File #1 not found}
						\if@bbllx\if@bblly
				   		\if@bburx\if@bbury
			      				\def\@p@sfile{#1}%
			      				\def\@p@sbbfile{#1}%
							\@decmprfalse
				  	   	\fi\fi\fi\fi
					\else\closein1
				    		\def\@p@sfile{\figurepath#1}%
				    		\def\@p@sbbfile{\figurepath#1}%
						\@decmprfalse
	                       		\fi%
			 	\else\closein1%
					\def\@p@sfile{#1}
					\def\@p@sbbfile{#1}
					\@decmprfalse
			 	\fi
			\else
				\def\@p@sfile{\figurepath#1}
				\def\@p@sbbfile{\figurepath#1.bb}
				\@decmprtrue
			\fi
		\else
			\def\@p@sfile{#1}
			\def\@p@sbbfile{#1.bb}
			\@decmprtrue
		\fi}
 
\def\@p@@sfile#1{\@p@@sfigure{#1}}
 
\def\@p@@sbbllx#1{
		\@bbllxtrue
		\dimen100=#1
		\edef\@p@sbbllx{\number\dimen100}
}
\def\@p@@sbblly#1{
		\@bbllytrue
		\dimen100=#1
		\edef\@p@sbblly{\number\dimen100}
}
\def\@p@@sbburx#1{
		\@bburxtrue
		\dimen100=#1
		\edef\@p@sbburx{\number\dimen100}
}
\def\@p@@sbbury#1{
		\@bburytrue
		\dimen100=#1
		\edef\@p@sbbury{\number\dimen100}
}
\def\@p@@sheight#1{
		\@heighttrue
		\dimen100=#1
   		\edef\@p@sheight{\number\dimen100}
}
\def\@p@@swidth#1{
		\@widthtrue
		\dimen100=#1
		\edef\@p@swidth{\number\dimen100}
}
\def\@p@@srheight#1{
		\@rheighttrue
		\dimen100=#1
		\edef\@p@srheight{\number\dimen100}
}
\def\@p@@srwidth#1{
		\@rwidthtrue
		\dimen100=#1
		\edef\@p@srwidth{\number\dimen100}
}
\def\@p@@sangle#1{
		\@angletrue
		\edef\@p@sangle{#1} 
}
\def\@p@@ssilent#1{
		\@verbosefalse
}
\def\@p@@sprolog#1{\@prologfiletrue\def\@prologfileval{#1}}
\def\@p@@spostlog#1{\@postlogfiletrue\def\@postlogfileval{#1}}
\def\@cs@name#1{\csname #1\endcsname}
\def\@setparms#1=#2,{\@cs@name{@p@@s#1}{#2}}
\def\ps@init@parms{
		\@bbllxfalse \@bbllyfalse
		\@bburxfalse \@bburyfalse
		\@heightfalse \@widthfalse
		\@rheightfalse \@rwidthfalse
		\def\@p@sbbllx{}\def\@p@sbblly{}
		\def\@p@sbburx{}\def\@p@sbbury{}
		\def\@p@sheight{}\def\@p@swidth{}
		\def\@p@srheight{}\def\@p@srwidth{}
		\def\@p@sangle{0}
		\def\@p@sfile{} \def\@p@sbbfile{}
		\def\@p@scost{10}
		\def\@sc{}
		\@prologfilefalse
		\@postlogfilefalse
		\@clipfalse
		\if@noisy
			\@verbosetrue
		\else
			\@verbosefalse
		\fi
}
\def\parse@ps@parms#1{
	 	\@psdo\@psfiga:=#1\do
		   {\expandafter\@setparms\@psfiga,}}
\newif\ifno@bb
\def\bb@missing{
	\if@verbose{
		\ps@typeout{psfig: searching \@p@sbbfile \space  for bounding box}
	}\fi
	\no@bbtrue
	\epsf@getbb{\@p@sbbfile}
        \ifno@bb \else \bb@cull\epsf@llx\epsf@lly\epsf@urx\epsf@ury\fi
}
\def\bb@cull#1#2#3#4{
	\dimen100=#1 bp\edef\@p@sbbllx{\number\dimen100}
	\dimen100=#2 bp\edef\@p@sbblly{\number\dimen100}
	\dimen100=#3 bp\edef\@p@sbburx{\number\dimen100}
	\dimen100=#4 bp\edef\@p@sbbury{\number\dimen100}
	\no@bbfalse
}
\newdimen\p@intvaluex
\newdimen\p@intvaluey
\def\rotate@#1#2{{\dimen0=#1 sp\dimen1=#2 sp
		  \global\p@intvaluex=\cosine\dimen0
		  \dimen3=\sine\dimen1
		  \global\advance\p@intvaluex by -\dimen3
		  \global\p@intvaluey=\sine\dimen0
		  \dimen3=\cosine\dimen1
		  \global\advance\p@intvaluey by \dimen3
		  }}
\def\compute@bb{
		\no@bbfalse
		\if@bbllx \else \no@bbtrue \fi
		\if@bblly \else \no@bbtrue \fi
		\if@bburx \else \no@bbtrue \fi
		\if@bbury \else \no@bbtrue \fi
		\ifno@bb \bb@missing \fi
		\ifno@bb \ps@typeout{FATAL ERROR: no bb supplied or found}
			\no-bb-error
		\fi
		\count203=\@p@sbburx
		\count204=\@p@sbbury
		\advance\count203 by -\@p@sbbllx
		\advance\count204 by -\@p@sbblly
		\edef\ps@bbw{\number\count203}
		\edef\ps@bbh{\number\count204}
		\if@angle
			\Sine{\@p@sangle}\Cosine{\@p@sangle}
	        	{\dimen100=\maxdimen\xdef\r@p@sbbllx{\number\dimen100}
					    \xdef\r@p@sbblly{\number\dimen100}
			                    \xdef\r@p@sbburx{-\number\dimen100}
					    \xdef\r@p@sbbury{-\number\dimen100}}
                        \def\minmaxtest{
			   \ifnum\number\p@intvaluex<\r@p@sbbllx
			      \xdef\r@p@sbbllx{\number\p@intvaluex}\fi
			   \ifnum\number\p@intvaluex>\r@p@sbburx
			      \xdef\r@p@sbburx{\number\p@intvaluex}\fi
			   \ifnum\number\p@intvaluey<\r@p@sbblly
			      \xdef\r@p@sbblly{\number\p@intvaluey}\fi
			   \ifnum\number\p@intvaluey>\r@p@sbbury
			      \xdef\r@p@sbbury{\number\p@intvaluey}\fi
			   }
			\rotate@{\@p@sbbllx}{\@p@sbblly}
			\minmaxtest
			\rotate@{\@p@sbbllx}{\@p@sbbury}
			\minmaxtest
			\rotate@{\@p@sbburx}{\@p@sbblly}
			\minmaxtest
			\rotate@{\@p@sbburx}{\@p@sbbury}
			\minmaxtest
			\edef\@p@sbbllx{\r@p@sbbllx}\edef\@p@sbblly{\r@p@sbblly}
			\edef\@p@sbburx{\r@p@sbburx}\edef\@p@sbbury{\r@p@sbbury}
		\fi
		\count203=\@p@sbburx
		\count204=\@p@sbbury
		\advance\count203 by -\@p@sbbllx
		\advance\count204 by -\@p@sbblly
		\edef\@bbw{\number\count203}
		\edef\@bbh{\number\count204}
}
\def\in@hundreds#1#2#3{\count240=#2 \count241=#3
		     \count100=\count240	
		     \divide\count100 by \count241
		     \count101=\count100
		     \multiply\count101 by \count241
		     \advance\count240 by -\count101
		     \multiply\count240 by 10
		     \count101=\count240	
		     \divide\count101 by \count241
		     \count102=\count101
		     \multiply\count102 by \count241
		     \advance\count240 by -\count102
		     \multiply\count240 by 10
		     \count102=\count240	
		     \divide\count102 by \count241
		     \count200=#1\count205=0
		     \count201=\count200
			\multiply\count201 by \count100
		 	\advance\count205 by \count201
		     \count201=\count200
			\divide\count201 by 10
			\multiply\count201 by \count101
			\advance\count205 by \count201
		     \count201=\count200
			\divide\count201 by 100
			\multiply\count201 by \count102
			\advance\count205 by \count201
		     \edef\@result{\number\count205}
}
\def\compute@wfromh{
		\in@hundreds{\@p@sheight}{\@bbw}{\@bbh}
		\edef\@p@swidth{\@result}
}
\def\compute@hfromw{
	        \in@hundreds{\@p@swidth}{\@bbh}{\@bbw}
		\edef\@p@sheight{\@result}
}
\def\compute@handw{
		\if@height
			\if@width
			\else
				\compute@wfromh
			\fi
		\else
			\if@width
				\compute@hfromw
			\else
				\edef\@p@sheight{\@bbh}
				\edef\@p@swidth{\@bbw}
			\fi
		\fi
}
\def\compute@resv{
		\if@rheight \else \edef\@p@srheight{\@p@sheight} \fi
		\if@rwidth \else \edef\@p@srwidth{\@p@swidth} \fi
}
\def\compute@sizes{
	\compute@bb
	\if@scalefirst\if@angle
	\if@width
	   \in@hundreds{\@p@swidth}{\@bbw}{\ps@bbw}
	   \edef\@p@swidth{\@result}
	\fi
	\if@height
	   \in@hundreds{\@p@sheight}{\@bbh}{\ps@bbh}
	   \edef\@p@sheight{\@result}
	\fi
	\fi\fi
	\compute@handw
	\compute@resv}
 
\def\psfig#1{\vbox {
	\ps@init@parms
	\parse@ps@parms{#1}
	\compute@sizes
	\ifnum\@p@scost<\@psdraft{
		\special{ps::[begin] 	\@p@swidth \space \@p@sheight \space
				\@p@sbbllx \space \@p@sbblly \space
				\@p@sbburx \space \@p@sbbury \space
				startTexFig \space }
		\if@angle
			\special {ps:: \@p@sangle \space rotate \space}
		\fi
		\if@clip{
			\if@verbose{
				\ps@typeout{(clip)}
			}\fi
			\special{ps:: doclip \space }
		}\fi
		\if@prologfile
		    \special{ps: plotfile \@prologfileval \space } \fi
		\if@decmpr{
			\if@verbose{
				\ps@typeout{psfig: including \@p@sfile.Z \space }
			}\fi
			\special{ps: plotfile "`zcat \@p@sfile.Z" \space }
		}\else{
			\if@verbose{
				\ps@typeout{psfig: including \@p@sfile \space }
			}\fi
			\special{ps: plotfile \@p@sfile \space }
		}\fi
		\if@postlogfile
		    \special{ps: plotfile \@postlogfileval \space } \fi
		\special{ps::[end] endTexFig \space }
		\vbox to \@p@srheight sp{
			\hbox to \@p@srwidth sp{
				\hss
			}
		\vss
		}
	}\else{
		\if@draftbox{
			\hbox{\frame{\vbox to \@p@srheight sp{
			\vss
			\hbox to \@p@srwidth sp{ \hss \@p@sfile \hss }
			\vss
			}}}
		}\else{
			\vbox to \@p@srheight sp{
			\vss
			\hbox to \@p@srwidth sp{\hss}
			\vss
			}
		}\fi

	}\fi
}}
\psfigRestoreAt
\let\@=\LaTeXAtSign

\newread\epsffilein    
\newif\ifepsffileok    
\newif\ifepsfbbfound   
\newif\ifepsfverbose   
\newif\ifepsfdraft     
\newdimen\epsfxsize    
\newdimen\epsfysize    
\newdimen\epsftsize    
\newdimen\epsfrsize    
\newdimen\epsftmp      
\newdimen\pspoints     
\pspoints=1bp          
\epsfxsize=0pt         
\epsfysize=0pt         
\def\epsfbox#1{\global\def\epsfllx{72}\global\def\epsflly{72}%
   \global\def\epsfurx{540}\global\def\epsfury{720}%
   \def\lbracket{[}\def\testit{#1}\ifx\testit\lbracket
   \let\next=\epsfgetlitbb\else\let\next=\epsfnormal\fi\next{#1}}%
\def\epsfgetlitbb#1#2 #3 #4 #5]#6{\epsfgrab #2 #3 #4 #5 .\\%
   \epsfsetgraph{#6}}%
\def\epsfnormal#1{\epsfgetbb{#1}\epsfsetgraph{#1}}%
\def\epsfgetbb#1{%
\openin\epsffilein=#1
\ifeof\epsffilein\errmessage{I couldn't open #1, will ignore it}\else
   {\epsffileoktrue \chardef\other=12
    \def\do##1{\catcode`##1=\other}\dospecials \catcode`\ =10
    \loop
       \read\epsffilein to \epsffileline
       \ifeof\epsffilein\epsffileokfalse\else
          \expandafter\epsfaux\epsffileline:. \\%
       \fi
   \ifepsffileok\repeat
   \ifepsfbbfound\else
    \ifepsfverbose\message{No bounding box comment in #1; using defaults}\fi\fi
   }\closein\epsffilein\fi}%
\def\epsfclipon{\def\epsfclipstring{ clip}}%
\def\epsfclipoff{\def\epsfclipstring{\ifepsfdraft\space clip\fi}}%
\epsfclipoff
\def\epsfsetgraph#1{%
   \epsfrsize=\epsfury\pspoints
   \advance\epsfrsize by-\epsflly\pspoints
   \epsftsize=\epsfurx\pspoints
   \advance\epsftsize by-\epsfllx\pspoints
   \epsfxsize\epsfsize\epsftsize\epsfrsize
   \ifnum\epsfxsize=0 \ifnum\epsfysize=0
      \epsfxsize=\epsftsize \epsfysize=\epsfrsize
      \epsfrsize=0pt
     \else\epsftmp=\epsftsize \divide\epsftmp\epsfrsize
       \epsfxsize=\epsfysize \multiply\epsfxsize\epsftmp
       \multiply\epsftmp\epsfrsize \advance\epsftsize-\epsftmp
       \epsftmp=\epsfysize
       \loop \advance\epsftsize\epsftsize \divide\epsftmp 2
       \ifnum\epsftmp>0
          \ifnum\epsftsize<\epsfrsize\else
             \advance\epsftsize-\epsfrsize \advance\epsfxsize\epsftmp \fi
       \repeat
       \epsfrsize=0pt
     \fi
   \else \ifnum\epsfysize=0
     \epsftmp=\epsfrsize \divide\epsftmp\epsftsize
     \epsfysize=\epsfxsize \multiply\epsfysize\epsftmp   
     \multiply\epsftmp\epsftsize \advance\epsfrsize-\epsftmp
     \epsftmp=\epsfxsize
     \loop \advance\epsfrsize\epsfrsize \divide\epsftmp 2
     \ifnum\epsftmp>0
        \ifnum\epsfrsize<\epsftsize\else
           \advance\epsfrsize-\epsftsize \advance\epsfysize\epsftmp \fi
     \repeat
     \epsfrsize=0pt
    \else
     \epsfrsize=\epsfysize
    \fi
   \fi
   \ifepsfverbose\message{#1: width=\the\epsfxsize, height=\the\epsfysize}\fi
   \epsftmp=10\epsfxsize \divide\epsftmp\pspoints
   \vbox to\epsfysize{\vfil\hbox to\epsfxsize{%
      \ifnum\epsfrsize=0\relax
        \includegraphics{\ifepsfdraft}%
      \else
        \epsfrsize=10\epsfysize \divide\epsfrsize\pspoints
        \includegraphics{\ifepsfdraft}%
      \fi
      \hfil}}%
\global\epsfxsize=0pt\global\epsfysize=0pt}%
{\catcode`\%=12 \global\let\epsfpercent=
\long\def\epsfaux#1#2:#3\\{\ifx#1\epsfpercent
   \def\testit{#2}\ifx\testit\epsfbblit
      \epsfgrab #3 . . . \\%
      \epsffileokfalse
      \global\epsfbbfoundtrue
   \fi\else\ifx#1\par\else\epsffileokfalse\fi\fi}%
\def\epsfempty{}%
\def\epsfgrab #1 #2 #3 #4 #5\\{%
\global\def\epsfllx{#1}\ifx\epsfllx\epsfempty
      \epsfgrab #2 #3 #4 #5 .\\\else
   \global\def\epsflly{#2}%
   \global\def\epsfurx{#3}\global\def\epsfury{#4}\fi}%
\def\epsfsize#1#2{\epsfxsize}
\let\epsffile=\epsfbox

\renewcommand{\baselinestretch}{1.5}
\renewcommand{\thefootnote}{\fnsymbol{footnote}}
\topmargin -0.25in
\textheight 8.5in
\oddsidemargin 0.25in
\textwidth 6.1in
\font\elevenrm=cmr9 scaled\magstep1      \let\elrm=\elevenrm
  \font\elevensl=cmsl9 scaled\magstephalf  \let\elsl=\elevensl
    \font\elevenbf=cmbx9 scaled\magstep1     \let\elbf=\elevenbf
      \font\elevenit=cmti9 scaled\magstephalf   \let\elit=\elevenit
 
\font\twelvebsl=cmbxsl10 scaled\magstep 1
   \let\bsl=\twelvebsl
\def\mbf#1{\hbox{${\twelvebsl #1}$}}
\font\tenbsl=cmbxsl10
 
\font\eightrm=cmr8 scaled\magstep1
 
\font\ninerm=cmr9 scaled\magstep1
\catcode`\@=11 
\makeatletter
\def\@seccntformat#1{\csname the#1\endcsname.\hskip 1em}
\renewcommand\thesubsection{\Alph{subsection}}
\makeatother
\pagestyle{plain}
\begin{document}

\thispagestyle{empty}
\begin{flushright}
{\footnotesize\renewcommand{\baselinestretch}{.75}
SLAC--PUB--7914\\
OUNP-98-06\\
August 1998\\
}
\end{flushright}
\begin{center}
 {\large \bf 
HEAVY QUARK MASS EFFECTS AND IMPROVED TESTS OF THE FLAVOUR INDEPENDENCE 
OF STRONG INTERACTIONS$^*$}

\vskip  1truecm
{\bf P.N. BURROWS}

University of Oxford,\\ 
Particle and Nuclear Physics,\\ 
Keble Rd., Oxford, OX1 3RH, UK\\
{\it E-mail: p.burrows1@physics.ox.ac.uk}   

\vskip .5truecm
\centerline{Representing}

\vskip .3truecm
{\bf The SLD Collaboration$^{**}$}\\
Stanford Linear Accelerator Center \\
Stanford University, Stanford, CA~94309
\end{center}
 
\vskip .7truecm
\normalsize
 

\vskip .1truecm
 
\centerline{\bf ABSTRACT }
 
{\small
\noindent
A review is given of latest results on tests of the flavour independence of
strong interactions. Heavy quark mass effects are evident in the data and
are now taken into account at next-to-leading order in QCD perturbation theory.
The strong-coupling ratios \alpb/\alpuds and \alpc/\alpuds are found to be
consistent with unity. Determinations of the 
$b$-quark mass \mbz are discussed.
}

\vskip .2truecm

\vfill
 
\noindent
{\it Presented at the XXIX International Conference on High Energy Physics,
Vancouver, B.C. Canada, July 23-29 1998:
Parallel session 3 - QCD + Jet Physics}

\vskip .3truecm

\noindent
$^*$Work supported by Department of Energy contract DE-AC03-76SF00515 (SLAC).

\eject

\section{Motivation}
In order for Quantum Chromodynamics (QCD) to be a gauge-invariant
renormalisable field theory it is required that 
the strong coupling between quarks ($q$) and gluons ($g$), $\alpha_s$, 
be independent of quark flavour.
This basic {\it ansatz} can be tested directly in \epa by 
measuring the strong coupling in
events of the type \ep \ra \qqg for specific quark flavours.
Whereas an absolute determination of $\alpha_s$ using such a technique is
limited, primarily by large theoretical uncertainties, to the 5\%-level of
precision~\cite{phil}, a  much more precise test of the flavour-independence
can be made from the ratio of the couplings for different quark flavours,
in which most experimental errors and theoretical uncertainties cancel. 
Precise measurements of this type have been made using data collected at the 
\z0 resonance by the SLC and LEP collaborations, and are reviewed here.

However, the emission of
gluon radiation in \bb events is expected 
to be modified relative to that
in $q_l\overline{q_l}$ ($q_l$ =$u$+$d$+$s$) 
events due to the large $b$-quark mass, and such an effect must be taken into 
account in the test of flavour independence of the strong coupling. 
Recently three groups have completed
calculations of 3-jet observables in \z0 \ra \bb events complete at 
next-to-leading order in QCD perturbation 
theory~\cite{rodrigo,aachen,nason,bilenky}.  
Comparison of the rates for \z0 \ra \bbg and \z0 \ra $q_l\overline{q_l}g$ 
hence allows measurement of the mass\footnote{Use of the
modified minimal subtraction renormalisation scheme~\cite{msbar}
is implied throughout.} of the $b$-quark at the \z0 scale, \mbz.
Finally, given recent excitement in the electroweak
sector concerning the values of the quantities $R_b$ and $A_b$, it seems
worthwhile to provide a complementary 
cross-check of the strong dynamics of the $b$-quark.

\section{Strategy}

The analysis strategy is, in principle, straightforward: one chooses
infra-red- and collinear-safe observables to measure in \bb, \cc 
and $q_l\overline{q_l}$ events, and
compares the results with the QCD predictions, taking mass effects
into account.
However, there are at least two possible approaches: 1) input the best
knowledge of the $b$-mass taken from independent measurements, \eg in the
$\Upsilon$ system, and test the flavour-independence of the strong
coupling; 2) {\it assume} the flavour-independence, and determine the 
$b$-mass from the data. One or the other of these approaches has been followed 
in the results reported here.

The basic experimental method is common to both cases. One first selects \ep 
$\rightarrow$
hadrons events, and applies a flavour-tagging algorithm to select samples of 
events of different primary quark flavour, represented by the indices $i$ and
$j$. One measures the 3-jet observable(s) $X$ in the different tagged-flavour
event samples, and, in order to cancel sources of common systematic error,
forms the {\it ratios} $X^i/X^j$. Before these can be compared with the
relevant QCD calculations, they must be corrected for: the effects of
detector acceptance and resolution; the bias of the flavour tag to select
preferentially 2-jet rather than 3-jet events; the flavour compositions; and
hadronisation effects. These corrections are very important, and
the associated uncertainties must be taken into account in the estimation
of the systematic errors~\cite{delphi,newopal,newsld}.

\section{Flavour Tagging}

The three new contributions reviewed here have used complementary methods for
the separation of events of different primary quark flavour, and the
systematic uncertainties are therefore at least partly uncorrelated.
All methods rely on the large decay multiplicity and long lifetime of $B$
hadrons as a discriminator of \bb against \cc and 
$q_l\overline{q_l}$ events.

DELPHI has employed a method based on the probability $P$ that all tracks in 
an event originate from the primary \ep interaction point
(IP)~\cite{delphi}; $q_l\overline{q_l}$ events tend to have a high probability
value, and \bb events a low one. Requiring $P\leq0.005$ ($P > 0.2$)
tags a \bb~($q_l\overline{q_l}$) event sample with efficiency ($\epsilon$) and 
purity ($\Pi$) of 55\% and 85\% (80\% and 80\%) respectively.

OPAL has employed a method based on the number $N_{sig}$ of 
`significant' tracks, \ie
those whose impact parameter deviates significantly from the IP~\cite{newopal};
$q_l\overline{q_l}$
events tend to have few such tracks, and \bb events several. 
Requiring $N_{sig} \geq 5$ ($N_{sig}=0$) tags a 
\bb~($q_l\overline{q_l}$) event sample with
$\epsilon$, $\Pi$ = 23\%, 96\% (35\%, 86\%) respectively. 

SLD has employed a method based on the mass, $M_{vtx}$, and momentum,
$P_{vtx}$, of secondary decay vertices reconstructed using the 300M-pixel
CCD vertex detector~\cite{newsld}.
$q_l\overline{q_l}$ events rarely contain reconstructed
secondary decay vertices, and these typically result from
strange particle decays and are of low mass.
Conversely, \bb events typically contain high-mass vertices.
Requiring $M_{vtx} > 1.8$ GeV/$c^2$ tags a \bb event sample with
$\epsilon$, $\Pi$ = 62\%, 90\% respectively. Requiring no vertex and $N_{sig}$
= 0 tags a  $q_l\overline{q_l}$
event sample of $\epsilon$, $\Pi$ = 56\%, 91\% respectively.

In addition, both OPAL and SLD have tagged a third sample enriched in
primary \cc events. OPAL has explicitly reconstructed \Dst candidates with
$\epsilon$, $\Pi$ = 2\%, 55\% respectively. The low efficiency is due
mainly to the intrinsically low \Dst decay branching ratios into convenient
tagging modes. SLD has utilised the correlation between vertex mass and
momentum to distinguish \cc from \bb events by requiring $M_{vtx} < 1.8$
GeV/$c^2$ and $P_{vtx} > 5$ GeV/$c$. This yields a tagged sample with
higher $\epsilon$, $\Pi$ = 19\%, 64\%, and, equally importantly, low bias
against 3-jet events.

\section{Test of Flavour Independence}

Having measured the ratios $X^i/X^j$ for event flavour samples $i$ and $j$,
and made the corrections discussed in Section~2, 
the flavour-independence of strong interactions can be tested via
\begin{equation}
{X^i\over{X^j}}\;=\;{A^i(X)\,\alp^i\;+\;B^i(X)\,(\alp^i)^2\over
{A^j(X)\,\alp^j\;+\;B^j(X)\,(\alp^j)^2}}
\label{eq:qcd}
\end{equation}
where $A$ and $B$ are the perturbatively-calculated 
leading- (LO) and next-to-leading order (NLO) coefficients,
respectively, for observable $X$. In the case of \bb~(and \cc)
events these coefficients depend on the quark mass, and, for
a given input mass value, can be
calculated~\cite{bilenky}. Eq.~\ref{eq:qcd} can then be solved to obtain the
ratio $\alp^i/\alp^j$.
This has been done in new contributions from DELPHI, OPAL and SLD, using
$b$-mass values of \mbz=$2.67\pm0.27$, 
$M_b=5.0\pm0.5$ and \mbz=$3.0\pm0.5$ GeV/$c^2$, respectively (see Section~5).
Using the Durham (D)                              
jet-finder with a jet resolution parameter value $y_c=0.02$,
DELPHI has measured the 3-jet-rate ratio $R_3^{bl}$ $\equiv$ $R_3^b/R_3^{uds}$. 
Using the calculations of Ref.~\cite{rodrigo} they obtained~\cite{delphi}:
\begin{equation}
{\alpha_s^b \over{\alpha_s^{uds}}}=1.007\pm0.005({\rm exp.})\pm
0.007({\rm frag.})\pm0.005({\rm theor.})
\label{eq:delphias}
\end{equation}
OPAL has measured ratios of the distributions of the event shape obervables 
$D_2$, $1-T$, $M_H$, $B_T$, $B_W$ and $C$. Using the calculations
of Ref.~\cite{nason} they obtained~\cite{newopal}:
\begin{equation}
{\alpha_s^b \over{\alpha_s^{uds}}}=0.988\pm0.005({\rm stat.})
\pm0.004({\rm syst.})\pm0.011({\rm theor.})
\label{eq:opalasb}
\end{equation}
\begin{equation}
{\alpha_s^c \over{\alpha_s^{uds}}}=1.002\pm0.017({\rm stat.})
\pm0.025({\rm syst.})\pm0.009({\rm theor.})
\label{eq:opalasc}
\end{equation}
SLD has measured ratios of the 3-jet rates, $R_3$, for a wide range of $y_c$
values, for jets defined using the E, E0, P, P0, Durham and Geneva (G)
iterative clustering schemes.
Using the calculations
of Ref.~\cite{aachen} they obtained~\cite{newsld}:
\begin{equation}
{\alpha_s^b \over{\alpha_s^{uds}}}=1.004\pm0.018({\rm stat.})
^{+0.026}_{-0.031}({\rm syst.})^{+0.018}_{-0.029}({\rm theor.})
\label{eq:sldasb}
\end{equation}
\begin{equation}
{\alpha_s^c \over{\alpha_s^{uds}}}=1.036\pm0.043({\rm stat.})
^{+0.041}_{-0.045}({\rm syst.})^{+0.020}_{-0.018}({\rm theor.})
\label{eq:sldasc}
\end{equation}

\begin{figure}[htbp]
\center
\vspace*{-6.3cm}
\includegraphics{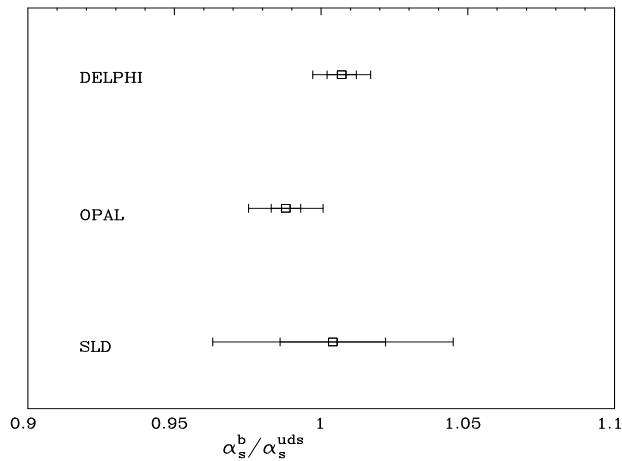}
\vskip 7.5truecm
\caption{Summary of new measurements of \alpb/\alpuds determined using quark
mass effects calculated at NLO.}
\label{fig:newas}
\end{figure}

All results are consistent with unity, \ie no flavour dependence of
the strong coupling. The SLD result has larger experimental errors as it 
is based on a data sample
comprising approximately 150k \z0 events, compared with the roughly 2.8M and
4.4M event samples used by DELPHI and OPAL, respectively.
The SLD analysis is currently being updated with the
additional 400k events collected between 1996 and 1998 with the new vertex
detector, and the resultant experimental precision is expected to be 
competitive with that achieved by the LEP experiments.
Interestingly the three collaborations quote rather different theoretical
uncertainties (Eqs.~\ref{eq:delphias}-\ref{eq:sldasc}). This may be due to
the facts that DELPHI used only one observable (see next section) and
implicitly assumed $\alp^c=\alp^{uds}$, that
different conventions were used for assigning the uncertainty, and that 
different procedures were used by OPAL and SLD for averaging 
over the ensemble of
observables. In order to compare results meaningfully and form a world average
it would be desirable to reach consensus on these issues.
The results for $\alp^b/\alp^{uds}$ are illustrated in Fig.~\ref{fig:newas}.
They are also in agreement with older results (not shown) 
incorporating mass effects at LO only~\cite{phil}.
\vskip 5truecm

\section{Heavy Quark Mass Effects}

The effects of a non-zero $b$-quark mass are clearly visible in the data.
For example, the SLD measured ratios $R_3^{bl}$ are shown in the bottom of 
Fig.~2(a).
For the E, E0, P and P0 algorithm cases $R_3^{bl}$
lies significantly
above unity (note that all the data points are highly correlated with one
another), whereas for the D and G algorithms $R_3^{bl}$ lies just below unity.
For a $b$-mass value of \mbb = $3.0\pm0.5$ GeV/$c^2$ the expected~\cite{aachen}
QCD values of $R_3^{bl}$ are shown as the short lines below the data points in
Fig.~2(a)
with the arrows pointing towards {\it lower} 
mass. The calculations are clearly in good agreement with the data.
Unfolding these data using Eq.~\ref{eq:qcd} yields the results for
\alpb/\alpuds shown at the bottom of 
Fig.~2(b)
and discussed in the previous section. 

\begin{figure}[htbp]
 \begin{center}
 \leavevmode
 \epsfxsize=4in
 \epsfysize=4in
 \epsfbox{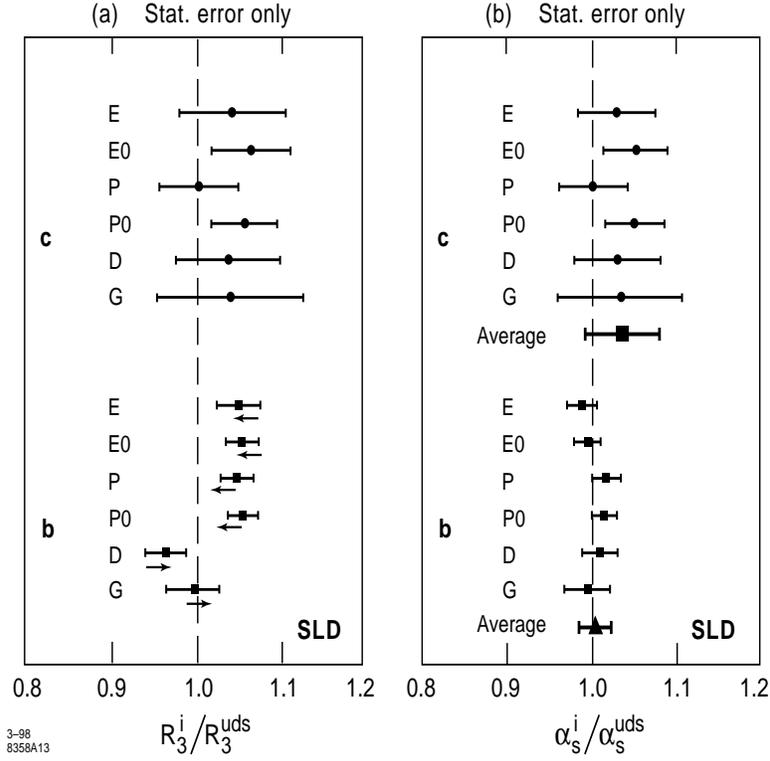} 
 \caption{$R_3^i/R_3^{uds}$ and $\alpha_s^i$/\alpuds measurements for $i=c,b$.}
 \end{center}
 \label{fig:rsld}
\end{figure}

\begin{figure}[htbp]
 \begin{center}
 \leavevmode
 \epsfxsize=3.5in
 \epsfysize=3.5in
 \epsfbox{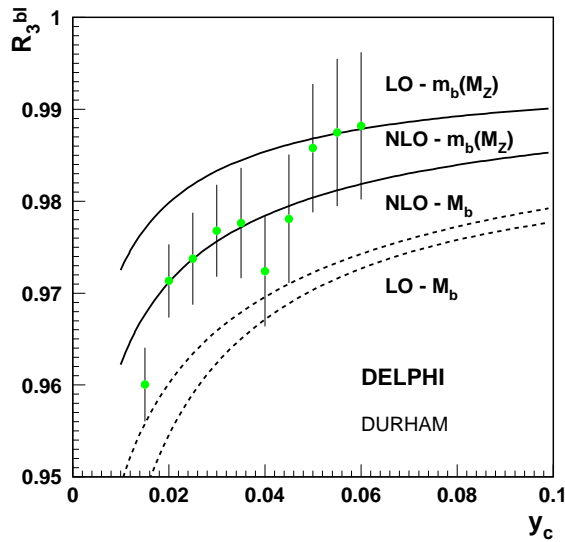} 
 \caption{$R_3^{bl}$ vs. $y_c$ compared with QCD calculations using $M_b$
 and \mbz (see text).}
 \label{fig:deldur}
 \end{center}
\end{figure}

One sees, therefore, a $b$-mass effect that is jet-algorithm dependent, 
with $R_3^{bl} \geq 1$ for
the JADE family of algorithms and $R_3^{bl} \leq 1$ for the D and G 
algorithms,
and which is reproduced beautifully by the NLO calculations. 
Qualitatively similar effects were found for the observables used 
in the OPAL study~\cite{newopal}. This can be
understood simply in terms of two competing physical origins.
1) The non-zero $b$-mass tends to cause a phase-space suppression of gluon
emission relative to the massless quark case. 2)
For a given kinematic configuration, 
the large $b$-mass tends to enhance the invariant mass of a local quark-gluon
pair relative to the massless quark case.
Since the JADE family of jet
algorithms is based on a clustering metric that is closely related to
invariant mass, for fixed $y_c$ the two partons are more likely to be resolved 
as separate jets when the quark is massive, implying $R_3^{bl} \geq 1$. 
By contrast, the
clustering metric used in the Durham and Geneva algorithms is less sensitive
to this kinematic effect, the phase-space suppression dominates, and
$R_3^{bl} \leq 1$. For increasing values of $y_c$ one expects both effects 1) 
and 2) to diminish in importance, and $R_3^{bl}$ \ra 1. This is indeed 
observed, as
illustrated in Fig.~\ref{fig:deldur} from DELPHI, which shows $R_3^{bl}(y_c)$
for the D case.

\begin{figure}[htbp]
 \begin{center}
 \leavevmode
 \epsfxsize=3.5in
 \epsfysize=3.5in
 \epsfbox{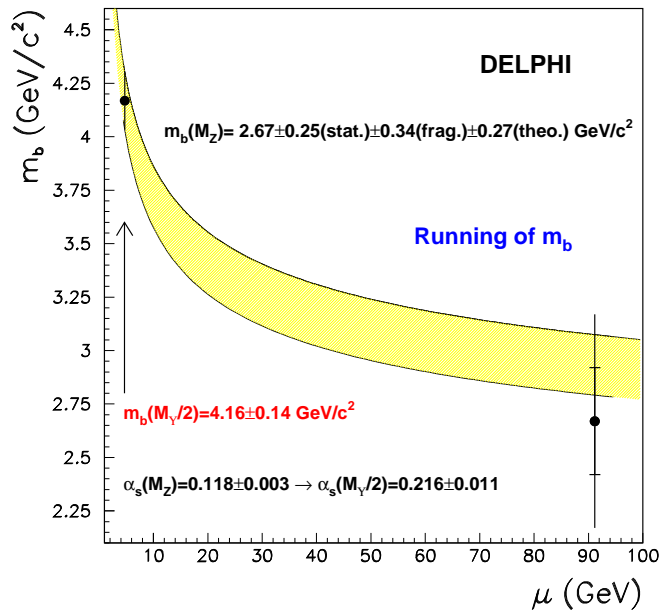} 
 \caption{The running of the $b$-mass with energy scale.}
 \label{fig:delrun}
 \end{center}
\end{figure}

Also shown in Fig.~\ref{fig:deldur} are curves representing the LO and NLO
QCD calculations~\cite{rodrigo} for two choices of the $b$-mass based on
$\Upsilon$ data:
the pole mass $M_b$ (=4.6 GeV/$c^2$)  
and the corresponding {\it running} mass \mbz~(=2.8 GeV/$c^2$). 
The NLO prediction with \mbz 
clearly provides the best description of the data, and
DELPHI has fitted it to the data point at $y_c$ = 0.02 by allowing the
$b$-mass \mbz to vary. They obtained~\cite{delphi}:
\begin{equation}
m_b(M_Z)= 2.67 \pm 0.25 ({\rm exp.})\pm0.34({\rm frag.})
\pm0.27 ({\rm theor.})\;{\rm GeV}/c^2
\label{eq:delphi}
\end{equation}
This result is illustrated in Fig.~\ref{fig:delrun}, where it is plotted at the
energy scale $\mu$ = $M_Z$ and compared with the value~\cite{delphi}
$m_b(M_{\Upsilon/2}) = 4.16\pm0.14$ GeV/$c^2$. The comparison yields:
\begin{equation}
m_b(M_Z)-m_b(M_{\Upsilon/2})\;\;=\;\;-1.49\pm0.52\;
{\rm GeV}/c^2 
\label{eq:delrun}
\end{equation}
which is consistent with the expected QCD running, shown by the band in
Fig.~\ref{fig:delrun}.

\begin{figure}[htbp]
 \begin{center}
 \leavevmode
 \epsfxsize=3.5in
 \epsfysize=3.5in
 \epsfbox{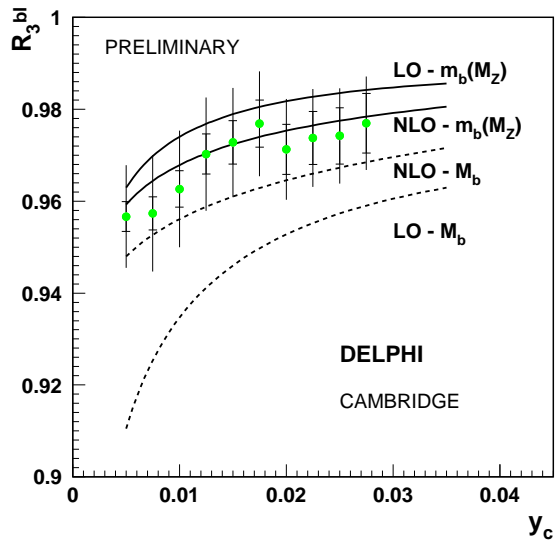} 
 \caption{$R_3^{bl}$ vs. $y_c$ compared with QCD calculations using $M_b$
 and \mbz (see text).}
 \label{fig:delcam}
 \end{center}
\end{figure}

DELPHI has updated its study recently~\cite{newdelphi} with an improved
flavour tag,
and has also employed the new Cambridge jet-finding algorithm~\cite{cambridge}. 
The corresponding dependence of $R_3^{bl}$ on $y_c$ is shown
in Fig.~\ref{fig:delcam}. In this case, although the NLO calculation with \mbz
describes the data best, the NLO and running mass effects do not seem as
pronounced as in the Durham case, and both the LO calculation with \mbz and the
NLO calculation with $M_b$ are also consistent with the data.

\begin{figure}[hbtp]
 \leavevmode
 \epsfxsize=6in
 \epsfysize=6in
 \epsfbox{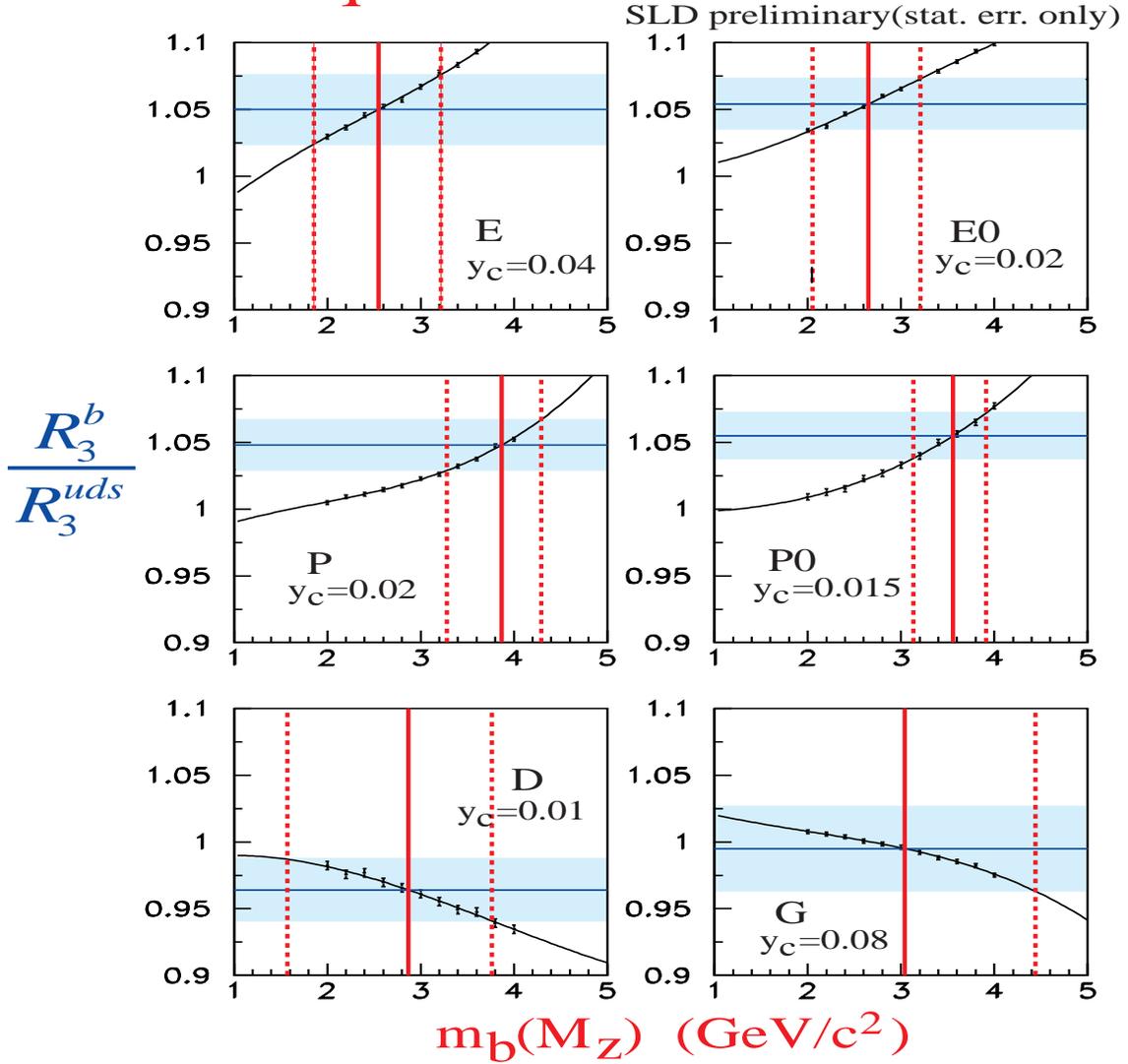} 
\caption{SLD $R_3^b/R_3^{uds}$ measurements compared with the \mbz-dependence
of the NLO calculation (see text).}
 \label{fig:sldbmass}
\end{figure}

The running $b$-mass has also been studied~\cite{pnb} using the SLD data 
shown in 
Fig.~2.
The calculations described in 
Ref.~\cite{aachen} have been performed, for the six jet algorithms and $y_c$
values used, for \mbz values in the range $2.0 \leq \mbz \leq 4.0$ GeV/$c^2$. 
The results are illustrated in Fig.~\ref{fig:sldbmass}, where the curves are
cubic polynomial interpolations between the calculations at discrete mass
values. It can be seen that the \mbz-dependence has a different form for
each algorithm, with a positive slope for the E, E0, P and P0 cases, and a 
negative slope for the D and G cases.
By comparing these curves with the SLD data, 
shown as horizontal bands, for each algorithm one can
read off the value of \mbz that is preferred, represented by the vertical
lines.  

It is worth reemphasising that the SLD results shown are highly correlated 
among the six jet algorithms since the same data sample was used and the
observables are intrinsically correlated. It is therefore interesting that
no single $b$-mass value fits the data corresponding to all six algorithms.
The r.m.s. scatter of the central values is $\Delta$\mbz = $\pm0.49$ GeV/$c^2$,
which is comparable with the DELPHI total error on \mbz using just the Durham
algorithm. The origin of this scatter is unclear, but it is certainly
plausible that it results from higher-order QCD effects that
are, by definition, not included in the NLO calculation, and which can 
\apriori be of different sign and magnitude for different obsrvables.
In addition, the considerable dependence~\cite{newsld} 
on jet algorithm of the size of the hadronisation uncertainties 
on $R_3^{bl}$ translates~\cite{pnb} into uncertainties on \mbz as large as
$^{+0.5}_{-1.7}$ GeV/$c^2$.

A weighted average over the results from the six jet algorithms yields a
preliminary result:
\begin{equation}                                                   
m_b(M_Z)=3.23^{+0.56}_{-0.72}({\rm stat.})^{+0.81}_{-1.28}{\rm (syst.})
^{+0.28}_{-1.05} ({\rm theor.}) \pm0.49 {\rm (r.m.s.)}\; {\rm GeV}/c^2     
\label{eq:sldmass}
\end{equation}
This is consistent with Eq.~\ref{eq:delphi}, and the r.m.s. has been
conservatively included as an additional uncertainty.
(For this reason the r.m.s. over the six  \alpb/\alpuds results discussed in the
previous section was included as an uncertainty on the SLD average value,
Eq.~\ref{eq:sldasb}). 
Finally, it should be noted that effects due to the $c$-quark mass are 
expected to be much smaller than those due to the $b$-mass, \ie at the level of
1\% or less on $R_3^c/R_3^{uds}$. Such effects are much smaller than the
current experimental errors, and are difficult to
observe; see 
Fig.~2
and Refs.~\cite{newopal,newsld}.

\section{Summary and Conclusions}

Recent results on tests of the flavour independence of strong interactions
and heavy quark mass effects from DELPHI, OPAL and SLD have been reviewed.
Three theoretical groups have recently completed 
calculations of 3-jet observables, including quark mass effects, 
complete at next-to-leading order in perturbative QCD.
These calculations have been used
enthusiastically by the experimentalists in the flavour-independence studies,
as well as to determine the $b$-quark mass at the \z0 scale.

The values of the strong coupling ratios,  
$\alpha_s^b/\alpha_s^{uds}$ and
$\alpha_s^c/\alpha_s^{uds}$,
extracted using the different NLO
calculations, are in good agreement with one another, with the previous
results incorporating mass effects only at LO, and with unity.
There is hence no evidence for
any anomalous flavour-dependent effects.
Though the measurements of \alpb/\alpuds are approaching the per cent level of 
precision, there are fewer measurements of \alpc/\alpuds, and these are
limited in precision currently to the 5\% level; more and better measurements
would be welcomed. It would also be desirable to achieve a consensus on the
treatment of experimental and theoretical systematic uncertainties by the
different collaborations, so that the data could be combined sensibly to
obtain meaningful world average values and errors.

The DELPHI and SLD data have been used to determine \mbz and consistent values 
are obtained. A strong jet-algorithm dependence of \mbz has been 
observed, as
well as large hadronisation uncertainties for some jet algorithms. 
DELPHI is
currently investigating use of the new Cambridge jet algorithm, and the recent
large
SLD data sample collected with the new vertex detector will be included in
an analysis optimised for the study of \mbz. 

\section*{Acknowledgements}
I thank my colleagues in the SLD Collaboration for their support, especially
D.~Muller for careful reading of this manuscript. I thank
A. Brandenburg and his collaborators for their heroic calculations and many
helpful discussions.

\vskip .4truecm
 
\vbox{\footnotesize\renewcommand{\baselinestretch}{1}\noindent
$^*$Work supported by Department of Energy
  contracts:
  DE-FG02-91ER40676 (BU),
  DE-FG03-91ER40618 (UCSB),
  DE-FG03-92ER40689 (UCSC),
  DE-FG03-93ER40788 (CSU),
  DE-FG02-91ER40672 (Colorado),
  DE-FG02-91ER40677 (Illinois),
  DE-AC03-76SF00098 (LBL),
  DE-FG02-92ER40715 (Massachusetts),
  DE-FC02-94ER40818 (MIT),
  DE-FG03-96ER40969 (Oregon),
  DE-AC03-76SF00515 (SLAC),
  DE-FG05-91ER40627 (Tennessee),
  DE-FG02-95ER40896 (Wisconsin),
  DE-FG02-92ER40704 (Yale);
  National Science Foundation grants:
  PHY-91-13428 (UCSC),
  PHY-89-21320 (Columbia),
  PHY-92-04239 (Cincinnati),
  PHY-95-10439 (Rutgers),
  PHY-88-19316 (Vanderbilt),
  PHY-92-03212 (Washington);
  The UK Particle Physics and Astronomy Research Council
  (Brunel, Oxford and RAL);
  The Istituto Nazionale di Fisica Nucleare of Italy
  (Bologna, Ferrara, Frascati, Pisa, Padova, Perugia);
  The Japan-US Cooperative Research Project on High Energy Physics
  (Nagoya, Tohoku);
  The Korea Research Foundation (Soongsil, 1997).}

\vskip 1truecm

\vfill
\eject

\vfill
\eject

\section*{$^{**}$List of Authors}


%
%
%
\begin{center}
\def\iADEL{$^{(1)}$}
\def\iAOMORI{$^{(2)}$}
\def\iBOLO{$^{(3)}$}
\def\iBRUN{$^{(4)}$}
\def\iBU{$^{(5)}$}
\def\iCINC{$^{(6)}$}
\def\iCOLO{$^{(7)}$}
\def\iCOLU{$^{(8)}$}
\def\iCSU{$^{(9)}$}
\def\iFERR{$^{(10)}$}
\def\iFRAS{$^{(11)}$}
\def\iILLI{$^{(12)}$}
\def\iLBL{$^{(13)}$}
\def\iLTU{$^{(14)}$}
\def\iMASS{$^{(15)}$}
\def\iMISSI{$^{(16)}$}
\def\iMIT{$^{(17)}$}
\def\iMOSCOW{$^{(18)}$}
\def\iNAGO{$^{(19)}$}
\def\iOREG{$^{(20)}$}
\def\iOXF{$^{(21)}$}
\def\iPADO{$^{(22)}$}
\def\iPERU{$^{(23)}$}
\def\iPISA{$^{(24)}$}
\def\iRAL{$^{(25)}$}
\def\iRUTG{$^{(26)}$}
\def\iSLAC{$^{(27)}$}
\def\iSOGA{$^{(28)}$}
\def\iSOONG{$^{(29)}$}
\def\iTENN{$^{(30)}$}
\def\iTOHO{$^{(31)}$}
\def\iUCSB{$^{(32)}$}
\def\iUCSC{$^{(33)}$}
\def\iVAND{$^{(34)}$}
\def\iWASH{$^{(35)}$}
\def\iWISC{$^{(36)}$}
\def\iYALE{$^{(37)}$}

  \baselineskip=.75\baselineskip  
\mbox{K. Abe\unskip,\iAOMORI}
\mbox{K.  Abe\unskip,\iNAGO}
\mbox{T. Abe\unskip,\iSLAC}
\mbox{I.Adam\unskip,\iSLAC}
\mbox{T.  Akagi\unskip,\iSLAC}
\mbox{N. J. Allen\unskip,\iBRUN}
\mbox{A. Arodzero\unskip,\iOREG}
\mbox{W.W. Ash\unskip,\iSLAC}
\mbox{D. Aston\unskip,\iSLAC}
\mbox{K.G. Baird\unskip,\iMASS}
\mbox{C. Baltay\unskip,\iYALE}
\mbox{H.R. Band\unskip,\iWISC}
\mbox{M.B. Barakat\unskip,\iLTU}
\mbox{O. Bardon\unskip,\iMIT}
\mbox{T.L. Barklow\unskip,\iSLAC}
\mbox{J.M. Bauer\unskip,\iMISSI}
\mbox{G. Bellodi\unskip,\iOXF}
\mbox{R. Ben-David\unskip,\iYALE}
\mbox{A.C. Benvenuti\unskip,\iBOLO}
\mbox{G.M. Bilei\unskip,\iPERU}
\mbox{D. Bisello\unskip,\iPADO}
\mbox{G. Blaylock\unskip,\iMASS}
\mbox{J.R. Bogart\unskip,\iSLAC}
\mbox{B. Bolen\unskip,\iMISSI}
\mbox{G.R. Bower\unskip,\iSLAC}
\mbox{J. E. Brau\unskip,\iOREG}
\mbox{M. Breidenbach\unskip,\iSLAC}
\mbox{W.M. Bugg\unskip,\iTENN}
\mbox{D. Burke\unskip,\iSLAC}
\mbox{T.H. Burnett\unskip,\iWASH}
\mbox{P.N. Burrows\unskip,\iOXF}
\mbox{A. Calcaterra\unskip,\iFRAS}
\mbox{D.O. Caldwell\unskip,\iUCSB}
\mbox{D. Calloway\unskip,\iSLAC}
\mbox{B. Camanzi\unskip,\iFERR}
\mbox{M. Carpinelli\unskip,\iPISA}
\mbox{R. Cassell\unskip,\iSLAC}
\mbox{R. Castaldi\unskip,\iPISA}
\mbox{A. Castro\unskip,\iPADO}
\mbox{M. Cavalli-Sforza\unskip,\iUCSC}
\mbox{A. Chou\unskip,\iSLAC}
\mbox{E. Church\unskip,\iWASH}
\mbox{H.O. Cohn\unskip,\iTENN}
\mbox{J.A. Coller\unskip,\iBU}
\mbox{M.R. Convery\unskip,\iSLAC}
\mbox{V. Cook\unskip,\iWASH}
\mbox{R. Cotton\unskip,\iBRUN}
\mbox{R.F. Cowan\unskip,\iMIT}
\mbox{D.G. Coyne\unskip,\iUCSC}
\mbox{G. Crawford\unskip,\iSLAC}
\mbox{C.J.S. Damerell\unskip,\iRAL}
\mbox{M. N. Danielson\unskip,\iCOLO}
\mbox{M. Daoudi\unskip,\iSLAC}
\mbox{N. de Groot\unskip,\iSLAC}
\mbox{R. Dell'Orso\unskip,\iPERU}
\mbox{P.J. Dervan\unskip,\iBRUN}
\mbox{R. de Sangro\unskip,\iFRAS}
\mbox{M. Dima\unskip,\iCSU}
\mbox{A. D'Oliveira\unskip,\iCINC}
\mbox{D.N. Dong\unskip,\iMIT}
\mbox{P.Y.C. Du\unskip,\iTENN}
\mbox{R. Dubois\unskip,\iSLAC}
\mbox{B.I. Eisenstein\unskip,\iILLI}
\mbox{V. Eschenburg\unskip,\iMISSI}
\mbox{E. Etzion\unskip,\iWISC}
\mbox{S. Fahey\unskip,\iCOLO}
\mbox{D. Falciai\unskip,\iFRAS}
\mbox{C. Fan\unskip,\iCOLO}
\mbox{J.P. Fernandez\unskip,\iUCSC}
\mbox{M.J. Fero\unskip,\iMIT}
\mbox{K.Flood\unskip,\iMASS}
\mbox{R. Frey\unskip,\iOREG}
\mbox{T. Gillman\unskip,\iRAL}
\mbox{G. Gladding\unskip,\iILLI}
\mbox{S. Gonzalez\unskip,\iMIT}
\mbox{E.L. Hart\unskip,\iTENN}
\mbox{J.L. Harton\unskip,\iCSU}
\mbox{A. Hasan\unskip,\iBRUN}
\mbox{K. Hasuko\unskip,\iTOHO}
\mbox{S. J. Hedges\unskip,\iBU}
\mbox{S.S. Hertzbach\unskip,\iMASS}
\mbox{M.D. Hildreth\unskip,\iSLAC}
\mbox{J. Huber\unskip,\iOREG}
\mbox{M.E. Huffer\unskip,\iSLAC}
\mbox{E.W. Hughes\unskip,\iSLAC}
\mbox{X.Huynh\unskip,\iSLAC}
\mbox{H. Hwang\unskip,\iOREG}
\mbox{M. Iwasaki\unskip,\iOREG}
\mbox{D. J. Jackson\unskip,\iRAL}
\mbox{P. Jacques\unskip,\iRUTG}
\mbox{J.A. Jaros\unskip,\iSLAC}
\mbox{Z.Y. Jiang\unskip,\iSLAC}
\mbox{A.S. Johnson\unskip,\iSLAC}
\mbox{J.R. Johnson\unskip,\iWISC}
\mbox{R.A. Johnson\unskip,\iCINC}
\mbox{T. Junk\unskip,\iSLAC}
\mbox{R. Kajikawa\unskip,\iNAGO}
\mbox{M. Kalelkar\unskip,\iRUTG}
\mbox{Y. Kamyshkov\unskip,\iTENN}
\mbox{H.J. Kang\unskip,\iRUTG}
\mbox{I. Karliner\unskip,\iILLI}
\mbox{H. Kawahara\unskip,\iSLAC}
\mbox{Y. D. Kim\unskip,\iSOGA}
\mbox{R. King\unskip,\iSLAC}
\mbox{M.E. King\unskip,\iSLAC}
\mbox{R.R. Kofler\unskip,\iMASS}
\mbox{N.M. Krishna\unskip,\iCOLO}
\mbox{R.S. Kroeger\unskip,\iMISSI}
\mbox{M. Langston\unskip,\iOREG}
\mbox{A. Lath\unskip,\iMIT}
\mbox{D.W.G. Leith\unskip,\iSLAC}
\mbox{V. Lia\unskip,\iMIT}
\mbox{C.-J. S. Lin\unskip,\iSLAC}
\mbox{X. Liu\unskip,\iUCSC}
\mbox{M.X. Liu\unskip,\iYALE}
\mbox{M. Loreti\unskip,\iPADO}
\mbox{A. Lu\unskip,\iUCSB}
\mbox{H.L. Lynch\unskip,\iSLAC}
\mbox{J. Ma\unskip,\iWASH}
\mbox{G. Mancinelli\unskip,\iRUTG}
\mbox{S. Manly\unskip,\iYALE}
\mbox{G. Mantovani\unskip,\iPERU}
\mbox{T.W. Markiewicz\unskip,\iSLAC}
\mbox{T. Maruyama\unskip,\iSLAC}
\mbox{H. Masuda\unskip,\iSLAC}
\mbox{E. Mazzucato\unskip,\iFERR}
\mbox{A.K. McKemey\unskip,\iBRUN}
\mbox{B.T. Meadows\unskip,\iCINC}
\mbox{G. Menegatti\unskip,\iFERR}
\mbox{R. Messner\unskip,\iSLAC}
\mbox{P.M. Mockett\unskip,\iWASH}
\mbox{K.C. Moffeit\unskip,\iSLAC}
\mbox{T.B. Moore\unskip,\iYALE}
\mbox{M.Morii\unskip,\iSLAC}
\mbox{D. Muller\unskip,\iSLAC}
\mbox{V.Murzin\unskip,\iMOSCOW}
\mbox{T. Nagamine\unskip,\iTOHO}
\mbox{S. Narita\unskip,\iTOHO}
\mbox{U. Nauenberg\unskip,\iCOLO}
\mbox{H. Neal\unskip,\iSLAC}
\mbox{M. Nussbaum\unskip,\iCINC}
\mbox{N.Oishi\unskip,\iNAGO}
\mbox{D. Onoprienko\unskip,\iTENN}
\mbox{L.S. Osborne\unskip,\iMIT}
\mbox{R.S. Panvini\unskip,\iVAND}
\mbox{H. Park\unskip,\iOREG}
\mbox{C. H. Park\unskip,\iSOONG}
\mbox{T.J. Pavel\unskip,\iSLAC}
\mbox{I. Peruzzi\unskip,\iFRAS}
\mbox{M. Piccolo\unskip,\iFRAS}
\mbox{L. Piemontese\unskip,\iFERR}
\mbox{E. Pieroni\unskip,\iPISA}
\mbox{K.T. Pitts\unskip,\iOREG}
\mbox{R.J. Plano\unskip,\iRUTG}
\mbox{R. Prepost\unskip,\iWISC}
\mbox{C.Y. Prescott\unskip,\iSLAC}
\mbox{G.D. Punkar\unskip,\iSLAC}
\mbox{J. Quigley\unskip,\iMIT}
\mbox{B.N. Ratcliff\unskip,\iSLAC}
\mbox{T.W. Reeves\unskip,\iVAND}
\mbox{J. Reidy\unskip,\iMISSI}
\mbox{P.L. Reinertsen\unskip,\iUCSC}
\mbox{P.E. Rensing\unskip,\iSLAC}
\mbox{L.S. Rochester\unskip,\iSLAC}
\mbox{P.C. Rowson\unskip,\iCOLU}
\mbox{J.J. Russell\unskip,\iSLAC}
\mbox{O.H. Saxton\unskip,\iSLAC}
\mbox{T. Schalk\unskip,\iUCSC}
\mbox{R.H. Schindler\unskip,\iSLAC}
\mbox{B.A. Schumm\unskip,\iUCSC}
\mbox{J. Schwiening\unskip,\iSLAC}
\mbox{S. Sen\unskip,\iYALE}
\mbox{V.V. Serbo\unskip,\iWISC}
\mbox{M.H. Shaevitz\unskip,\iCOLU}
\mbox{J.T. Shank\unskip,\iBU}
\mbox{G. Shapiro\unskip,\iLBL}
\mbox{D.J. Sherden\unskip,\iSLAC}
\mbox{K. D. Shmakov\unskip,\iTENN}
\mbox{C. Simopoulos\unskip,\iSLAC}
\mbox{N.B. Sinev\unskip,\iOREG}
\mbox{S.R. Smith\unskip,\iSLAC}
\mbox{M. B. Smy\unskip,\iCSU}
\mbox{J.A. Snyder\unskip,\iYALE}
\mbox{H. Staengle\unskip,\iCSU}
\mbox{A. Stahl\unskip,\iSLAC}
\mbox{P. Stamer\unskip,\iRUTG}
\mbox{R. Steiner\unskip,\iADEL}
\mbox{H. Steiner\unskip,\iLBL}
\mbox{M.G. Strauss\unskip,\iMASS}
\mbox{D. Su\unskip,\iSLAC}
\mbox{F. Suekane\unskip,\iTOHO}
\mbox{A. Sugiyama\unskip,\iNAGO}
\mbox{S. Suzuki\unskip,\iNAGO}
\mbox{M. Swartz\unskip,\iSLAC}
\mbox{A. Szumilo\unskip,\iWASH}
\mbox{T. Takahashi\unskip,\iSLAC}
\mbox{F.E. Taylor\unskip,\iMIT}
\mbox{J. Thom\unskip,\iSLAC}
\mbox{E. Torrence\unskip,\iMIT}
\mbox{N. K. Toumbas\unskip,\iSLAC}
\mbox{A.I. Trandafir\unskip,\iMASS}
\mbox{J.D. Turk\unskip,\iYALE}
\mbox{T. Usher\unskip,\iSLAC}
\mbox{C. Vannini\unskip,\iPISA}
\mbox{J. Va'vra\unskip,\iSLAC}
\mbox{E. Vella\unskip,\iSLAC}
\mbox{J.P. Venuti\unskip,\iVAND}
\mbox{R. Verdier\unskip,\iMIT}
\mbox{P.G. Verdini\unskip,\iPISA}
\mbox{S.R. Wagner\unskip,\iSLAC}
\mbox{D. L. Wagner\unskip,\iCOLO}
\mbox{A.P. Waite\unskip,\iSLAC}
\mbox{Walston, S.\unskip,\iOREG}
\mbox{J.Wang\unskip,\iSLAC}
\mbox{C. Ward\unskip,\iBRUN}
\mbox{S.J. Watts\unskip,\iBRUN}
\mbox{A.W. Weidemann\unskip,\iTENN}
\mbox{E. R. Weiss\unskip,\iWASH}
\mbox{J.S. Whitaker\unskip,\iBU}
\mbox{S.L. White\unskip,\iTENN}
\mbox{F.J. Wickens\unskip,\iRAL}
\mbox{B. Williams\unskip,\iCOLO}
\mbox{D.C. Williams\unskip,\iMIT}
\mbox{S.H. Williams\unskip,\iSLAC}
\mbox{S. Willocq\unskip,\iSLAC}
\mbox{R.J. Wilson\unskip,\iCSU}
\mbox{W.J. Wisniewski\unskip,\iSLAC}
\mbox{J. L. Wittlin\unskip,\iMASS}
\mbox{M. Woods\unskip,\iSLAC}
\mbox{G.B. Word\unskip,\iVAND}
\mbox{T.R. Wright\unskip,\iWISC}
\mbox{J. Wyss\unskip,\iPADO}
\mbox{R.K. Yamamoto\unskip,\iMIT}
\mbox{J.M. Yamartino\unskip,\iMIT}
\mbox{X. Yang\unskip,\iOREG}
\mbox{J. Yashima\unskip,\iTOHO}
\mbox{S.J. Yellin\unskip,\iUCSB}
\mbox{C.C. Young\unskip,\iSLAC}
\mbox{H. Yuta\unskip,\iAOMORI}
\mbox{G. Zapalac\unskip,\iWISC}
\mbox{R.W. Zdarko\unskip,\iSLAC}
\mbox{J. Zhou\unskip.\iOREG}

\it
  \vskip \baselineskip                   
  \centerline{(The SLD Collaboration)}   
  \vskip \baselineskip        
  \baselineskip=.75\baselineskip   
\iADEL
  Adelphi University,
  South Avenue-   Garden City,NY 11530, \break
\iAOMORI
  Aomori University,
  2-3-1 Kohata, Aomori City, 030 Japan, \break
\iBOLO
  INFN Sezione di Bologna,
  Via Irnerio 46    I-40126 Bologna  (Italy), \break
\iBRUN
  Brunel University,
  Uxbridge, Middlesex - UB8 3PH United Kingdom, \break
\iBU
  Boston University,
  590 Commonwealth Ave. - Boston,MA 02215, \break
\iCINC
  University of Cincinnati,
  Cincinnati,OH 45221, \break
\iCOLO
  University of Colorado,
  Campus Box 390 - Boulder,CO 80309, \break
\iCOLU
  Columbia University,
  Nevis Laboratories  P.O.Box 137 - Irvington,NY 10533, \break
\iCSU
  Colorado State University,
  Ft. Collins,CO 80523, \break
\iFERR
  INFN Sezione di Ferrara,
  Via Paradiso,12 - I-44100 Ferrara (Italy), \break
\iFRAS
  Lab. Nazionali di Frascati,
  Casella Postale 13   I-00044 Frascati (Italy), \break
\iILLI
  University of Illinois,
  1110 West Green St.  Urbana,IL 61801, \break
\iLBL
  Lawrence Berkeley Laboratory,
  Dept.of Physics 50B-5211 University of California-  Berkeley,CA 94720, \break
\iLTU
  Louisiana Technical University,
  , \break
\iMASS
  University of Massachusetts,
  Amherst,MA 01003, \break
\iMISSI
  University of Mississippi,
  University,MS 38677, \break
\iMIT
  Massachusetts Institute of Technology,
  77 Massachussetts Avenue  Cambridge,MA 02139, \break
\iMOSCOW
  Moscow State University,
  Institute of Nuclear Physics  119899 Moscow  Russia, \break
\iNAGO
  Nagoya University,
  Nagoya 464 Japan, \break
\iOREG
  University of Oregon,
  Department of Physics  Eugene,OR 97403, \break
\iOXF
  Oxford University,
  Oxford, OX1 3RH, United Kingdom, \break
\iPADO
  Universita di Padova,
  Via F. Marzolo,8   I-35100 Padova (Italy), \break
\iPERU
  Universita di Perugia, Sezione INFN,
  Via A. Pascoli  I-06100 Perugia (Italy), \break
\iPISA
  INFN, Sezione di Pisa,
  Via Livornese,582/AS  Piero a Grado  I-56010 Pisa (Italy), \break
\iRAL
  Rutherford Appleton Laboratory,
  Chiton,Didcot - Oxon OX11 0QX United Kingdom, \break
\iRUTG
  Rutgers University,
  Serin Physics Labs  Piscataway,NJ 08855-0849, \break
\iSLAC
  Stanford Linear Accelerator Center,
  2575 Sand Hill Road  Menlo Park,CA 94025, \break
\iSOGA
  Sogang University,
  Ricci Hall  Seoul, Korea, \break
\iSOONG
  Soongsil University,
  Dongjakgu Sangdo 5 dong 1-1    Seoul, Korea 156-743, \break
\iTENN
  University of Tennessee,
  401 A.H. Nielsen Physics Blg.  -  Knoxville,Tennessee 37996-1200, \break
\iTOHO
  Tohoku University,
  Bubble Chamber Lab. - Aramaki - Sendai 980 (Japan), \break
\iUCSB
  U.C. Santa Barbara,
  3019 Broida Hall  Santa Barbara,CA 93106, \break
\iUCSC
  U.C. Santa Cruz,
  Santa Cruz,CA 95064, \break
\iVAND
  Vanderbilt University,
  Stevenson Center,Room 5333  P.O.Box 1807,Station B  Nashville,TN 37235,
\break
\iWASH
  University of Washington,
  Seattle,WA 98105, \break
\iWISC
  University of Wisconsin,
  1150 University Avenue  Madison,WS 53706, \break
\iYALE
  Yale University,
  5th Floor Gibbs Lab. - P.O.Box 208121 - New Haven,CT 06520-8121. \break

\rm
%

\end{center}


\end{document}